\documentclass[a4paper,11pt,showkeys]{revtex4}
\usepackage[english]{babel}

\usepackage{latexsym}
\usepackage{amssymb}
\usepackage{amsfonts}
\usepackage{eucal}
\usepackage{mathrsfs}

\usepackage{graphicx}
\usepackage{psfrag}
\usepackage{url}

\begin{document}

\title{ Elasticity and metastability limit in supercooled liquids: a lattice model}

\author{ Alessandro Attanasi$^{*\dagger}$, Andrea Cavagna$^{\ddagger}$, and 
Jos\'e Lorenzana$^{\ddagger}$}

\affiliation{
$^{\dagger}$ Dipartimento di Fisica, Universit\`a di Roma ``La Sapienza'',
P.le Aldo Moro 2, 00185 Roma,  Italy
\\
$^{\ddagger}$Centre for Statistical Mechanics and Complexity--INFM and ISC--CNR, via dei Taurini $19$, $00185$ Roma, Italy}

\begin{abstract}
We present Monte Carlo simulations on a lattice system that displays a first 
order phase transition between a disordered phase (liquid) and an ordered phase
(crystal). The model is augmented by an interaction that simulates
the effect of elasticity in continuum models.   
The temperature range of stability of the liquid phase is
strongly increased in the presence of the elastic interaction. We
discuss the consequences of this result for the existence of a kinetic
spinodal in real systems.  
\end{abstract}


\maketitle

\section{Introduction}
When a liquid is cooled below its freezing point without forming a crystal, 
it enters a metastable equilibrium phase known as supercooled. A supercooled 
liquid is squeezed in an uncomfortable region in the time domain: 
if we are too fast in measuring 
its properties, the system cannot thermalize and it behaves as an 
off-equilibrium glass; while, if we are too slow, the system has 
the time to nucleate the solid, and we obtain an off-equilibrium
polycrystal. If at a certain 
temperature the relaxation time of the liquid $\tau_{\scriptscriptstyle R}$ 
exceeds the nucleation time of the crystal $\tau_{\scriptscriptstyle N}$, no 
equilibrium measurements can be performed on the liquid state and the 
supercooled phase does not exist anymore. Such a temperature is called kinetic 
spinodal $T_{sp}$ and it marks the metastability limit\cite{kauzmann48}.
Recently we have argued that the viscoelastic response of the supercooled 
liquid is the main mechanism determining whether 
or not a metastable limit is present\cite{cav05}.  The central 
idea is that on time scales shorter than $\tau_{\scriptscriptstyle R}$, 
supercooled liquids exhibit a solid-like response to strains. So, the 
thermodynamic drive for crystal nucleation gets depressed by an elastic 
contribution, which is relaxed with the passing of time. 

In order to partially test this idea we study here a lattice model,
without quenched disorder, that has a phenomenology similar to supercooled 
liquids (first order phase transition, metastability, glassy behaviour).
The advantage of using a lattice model rather
than a realistic structural supercooled liquid is that one can simulate for 
longer times and larger systems. 
We add in the model an {\em ad hoc} interaction 
which, we argue, simulates the effects of the elastic interaction as in a 
solid. Our proposal is to study  in a controlled manner 
how the elastic interactions affect the kinetic spinodal, neglecting, for 
simplicity, the relaxation of this interaction.

\section{The lattice model}
\begin{figure}[ht]
\includegraphics[clip=true,width=8cm,angle=0]{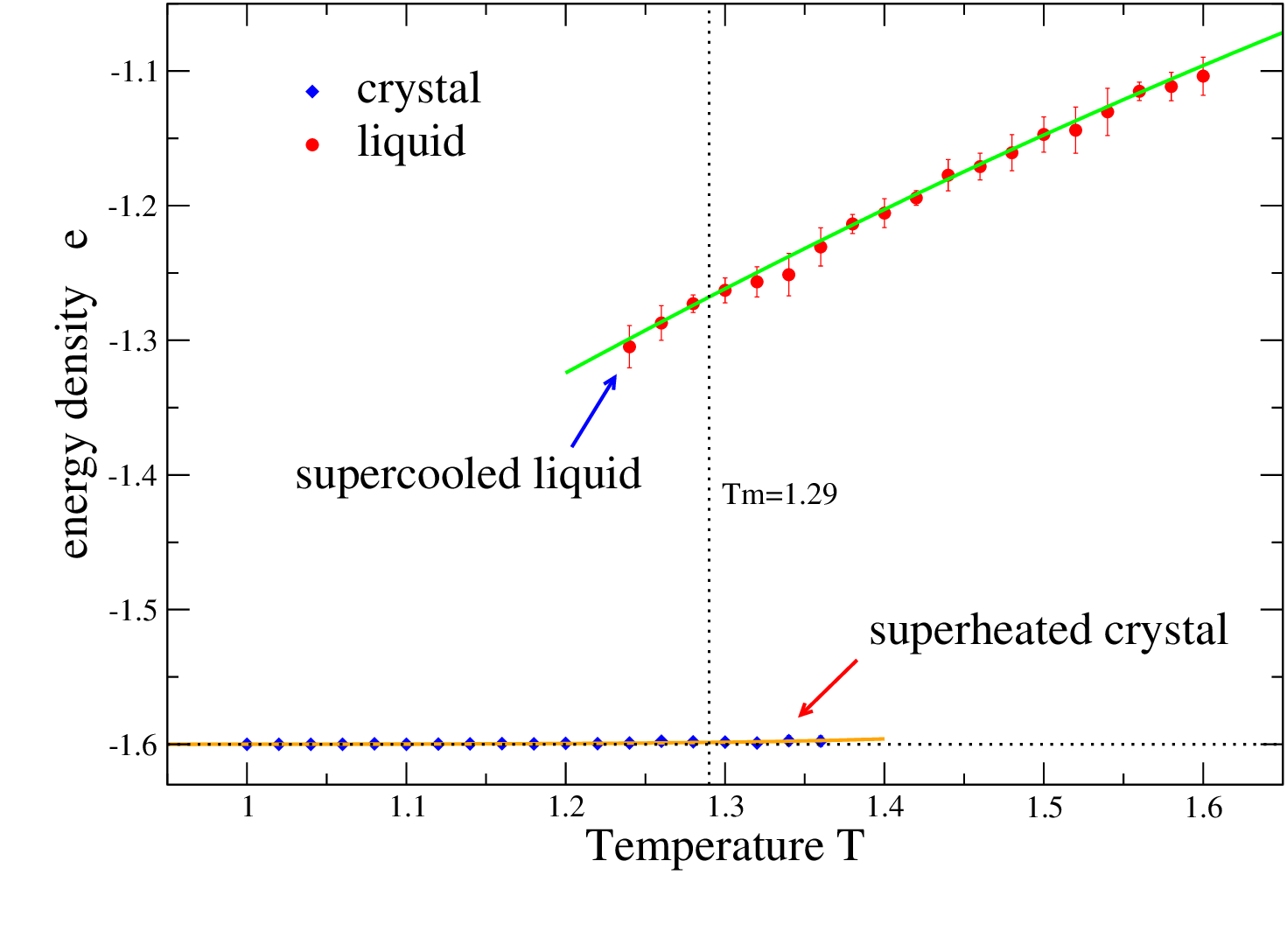}
\caption{\label{equ}
Equilibrium energy density of the liquid and the crystal as 
function of the temperature.}
\end{figure}

The lattice model, hereafter CTLS, without the elastic effects has been 
studied in Ref.~\cite{cavagna03}. The Hamiltonian is:
\begin{equation}
H_{{\scriptscriptstyle CTLS}}=J\sum_{i=1}^N (1+\sigma_i)f_i\;; 
\end{equation}
where $\sigma_i=\pm 1$ are spins on a 2-dimensional lattice and $f_i=\sigma_i^{\scriptscriptstyle N}\sigma_i^{\scriptscriptstyle S}\sigma_i^{\scriptscriptstyle E}\sigma_i^{\scriptscriptstyle W}$ is the product of the first neighbors of the
spin $\sigma_i$. We fix the units of energies and temperatures by
setting $J\equiv1$ and $k_B\equiv1$.  
The following results, in the absence of elastic effects,
 were found in  Ref.~\cite{cavagna03} and reproduced in the present study. 
The model present a first-order phase transition between 
a disordered phase (liquid) and an ordered one (crystal) at the melting 
temperature $T_m=1.29$ (Fig.~\ref{equ}). 

\begin{figure}[ht]
\includegraphics[clip=true,width=8cm,angle=0]{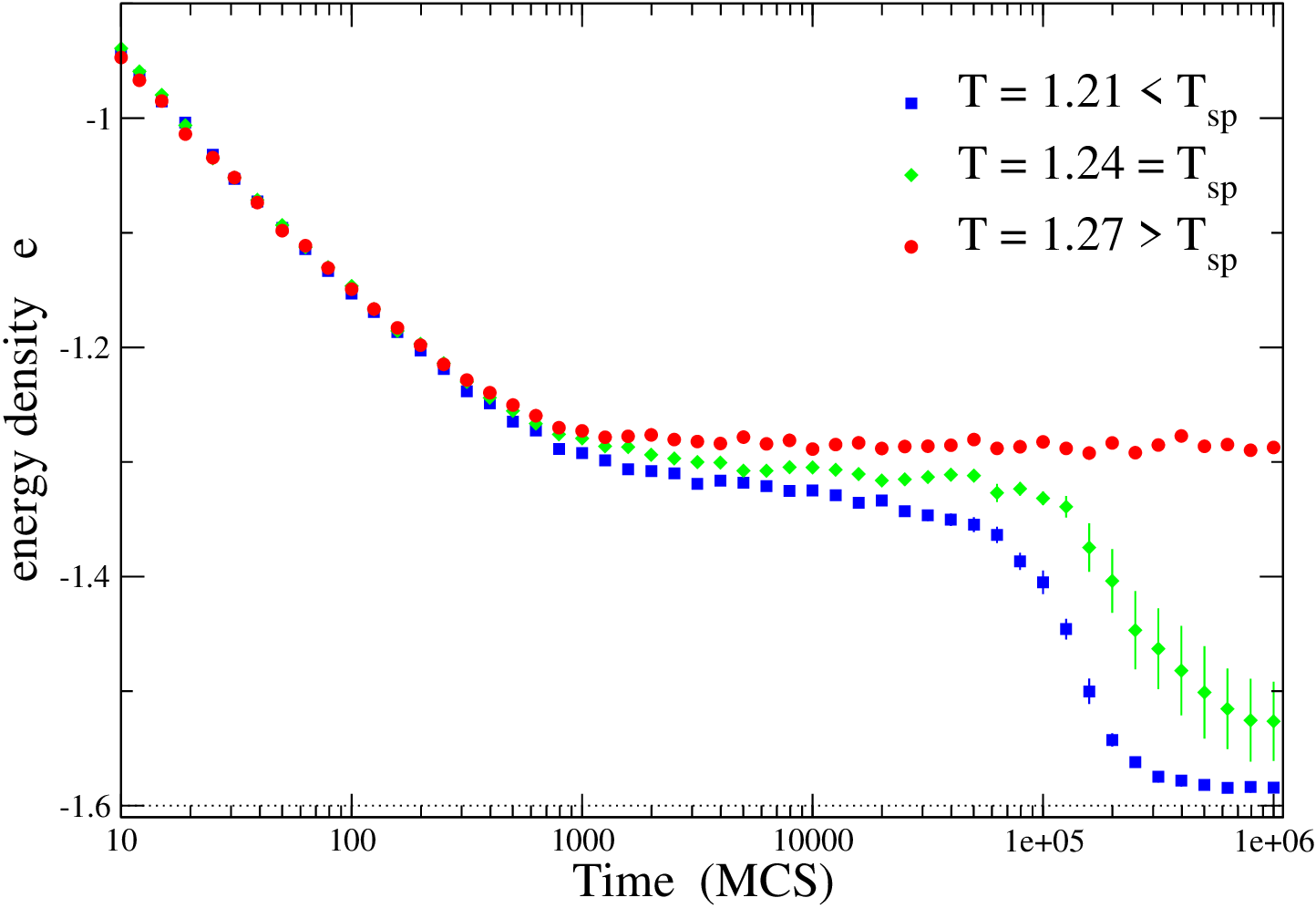}
\caption{\label{que}Quench experiments: energy density as function of time at 
three different temperatures for $p=0$.}
\end{figure}

The system has a well defined kinetic spinodal at temperature $T_{sp}$. This 
is illustrated in Fig.~\ref{que} where we show numerical quench experiments 
at different temperatures. If we quench the system at a temperature $T>T_{sp}$ 
we see that it remains trapped in a metastable liquid state for a time much 
longer than the relaxation time. 
Finally for $T<T_{sp}$ we don't observe any plateau and the 
system goes down towards the ground state. Below $T_{sp}$
the liquid phase cannot be equilibrated and therefore it becomes ill-defined. 

Before introducing the elastic effects in this model we have to define a  
``volume'' for the liquid phase and one for the crystal phase, and a 
``pressure''. We can write a new Hamiltonian for the system:
\begin{equation}
H=H_{{\scriptscriptstyle CTLS}}+pV\;,
\end{equation}
where $p$ is the pressure and $V$ is the total volume. 
We define an 
{\it ad hoc} variable $V$ which plays the role of volume: let us assume that the
volume per site for  the crystal and the liquid phase 
($\nu_{\scriptscriptstyle C}$, $\nu_{\scriptscriptstyle L}$ respectively) are different as  
in a real crystal-liquid transition
in the continuum. The total volume then can be put as:

\begin{eqnarray}
V=\nu_{\scriptscriptstyle L}N_{\scriptscriptstyle
  L}+\nu_{\scriptscriptstyle C}N_{\scriptscriptstyle
  C}=N\nu_{\scriptscriptstyle L}+N_{\scriptscriptstyle C} \Delta \nu\,.
\end{eqnarray}
here $N_{\scriptscriptstyle L}$ ($N_{\scriptscriptstyle C}$)  is the
number of sites in the liquid (crystal) phase and  $\Delta\nu=\nu_{\scriptscriptstyle C}-\nu_{\scriptscriptstyle L}$.
This defines the total volume as function of given quantities
$(\nu_{\scriptscriptstyle L},\nu_{\scriptscriptstyle C},N)$ and of the
total number of the spins that are in the crystalline phase.
To
determine the latter we define a local order parameter $m_i$ in such a
way that takes the value $m_i=1$ for every site in the crystalline
phase \cite{note} and we define the total number of crystalline sites as 
$N_C=\sum_{i=1}^N m_i$ \cite{note2}. We fix
our units of volume by setting  $\Delta \nu\equiv1$. In this units the
pressure is measure in units of $J/\Delta\nu$.

With these definitions the Hamiltonian takes the form:
\begin{eqnarray}
H=H_{\scriptscriptstyle {CTLS}}+p\Delta\nu N_C+p\nu_{\scriptscriptstyle L}N\,.
\end{eqnarray}   

\begin{figure}[ht]
\begin{center}
\includegraphics[clip=true,width=8cm]{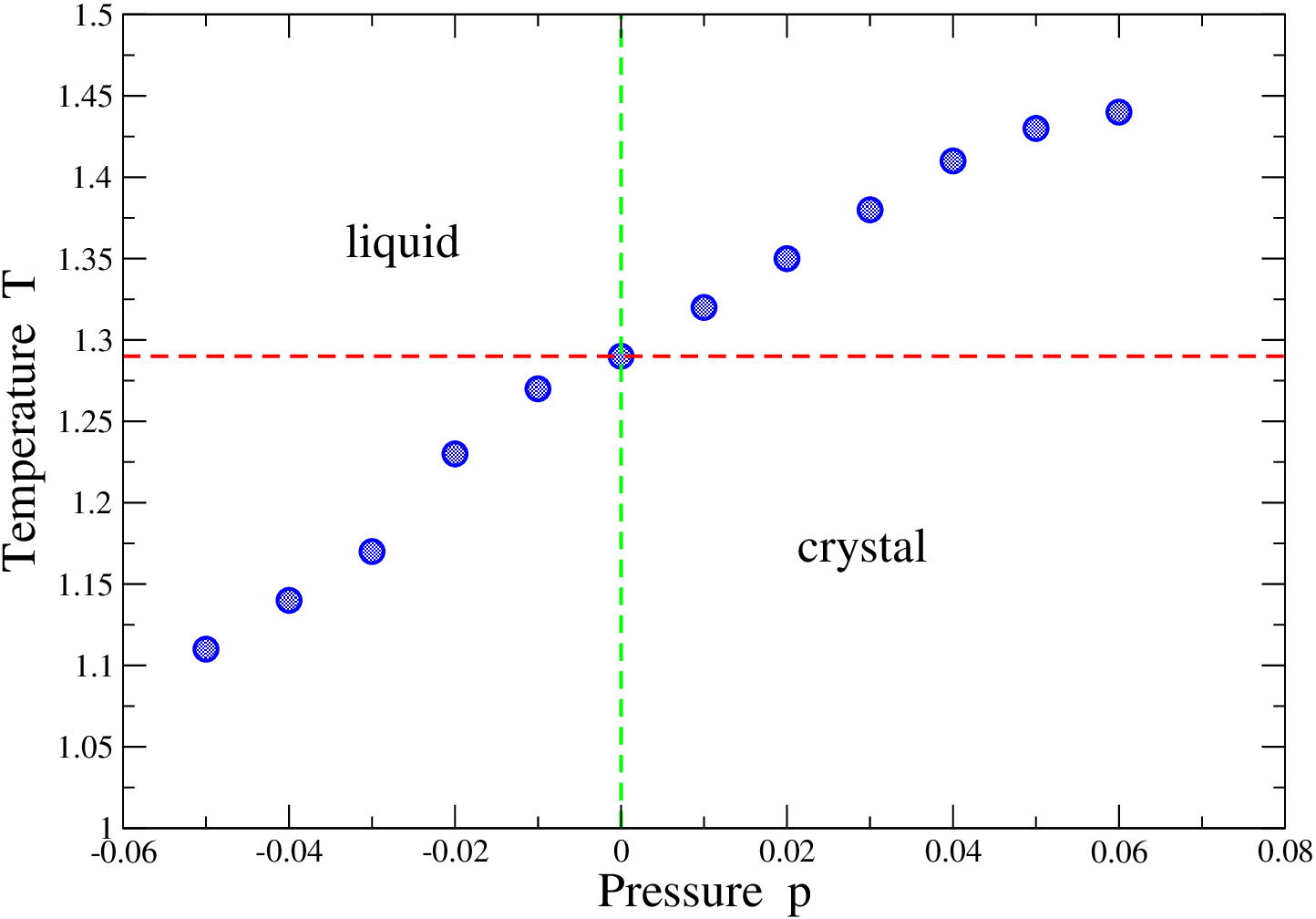}
\end{center}
\caption{\label{pdt}
Phase diagram: temperature vs pressure. Negative pressures correspond
to a strained system.}
\end{figure}

In Fig.~\ref{pdt} we show the 
phase diagram in the $p-V$ plane. We see that one obtains the expected 
behavior, indeed as the pressure increases the crystal phase, which we defined 
with the smallest specific volume, is favored.  

Sudden compression (crunching) experiments are particularly interesting because 
they are very similar to  quenching experiments but using the 
pressure as control variable. We can now monitor both  the value of energy 
density and the crystal volume. 
  
\begin{figure}[ht]
\begin{center}
\includegraphics[clip=true,width=11cm]{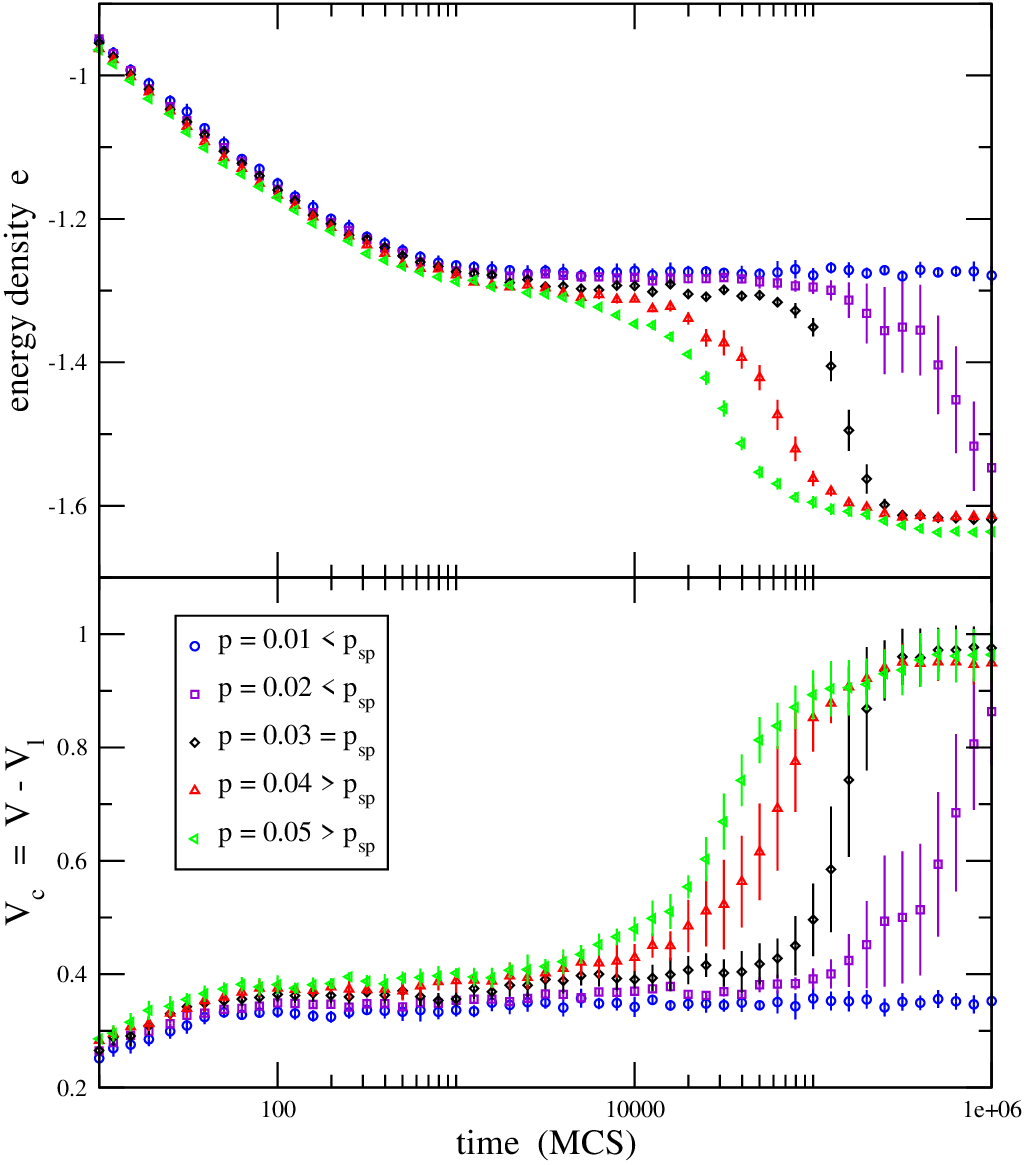}
\end{center}
\caption{\label{crunch}
Crunching experiments; top: energy density vs 
time, bottom: crystal volume vs time.
}
\end{figure}

As shown in 
Fig.~\ref{crunch} one can define a spinodal pressure $p_{\scriptscriptstyle sp}=0.03$. 
As long as we crunch the system at a pressure $p<p_{sp}$,
it remains trapped in the metastable liquid for a time much 
longer than the relaxation time. On the contrary, for $p>p_{sp}$ we 
do not observe any plateau: the system goes steadily (but slowly) to the ground state. 
Above $p_{sp}$ the liquid phase cannot be equilibrated and therefore it becomes ill-defined.
The same behaviour is displayed by the total crystal volume (Fig.~\ref{crunch}).

\section{Introducing elasticity in the model}
At low temperatures and for not too long times the supercooled liquid 
will show an elastic response to strains.  
Since the crystal has a different specific volume than the surrounding
liquid phase a relatively fast nucleation process will produce strains
in the liquid. The interplay between crystal nucleation and elastic stress relaxation is expressed by a self-consistent
equation for the nucleation time $\tau_{\mathrm N}$ \cite{cav05}:
\begin{eqnarray}
T\log \tau_{\mathrm N} = \frac{\sigma^3}{\left[ \delta G(T) - E_{\mathrm elastic}(f(\tau_{\mathrm N})/\tau_{\mathrm R})\right]^2} \ ,
\end{eqnarray}
where $\sigma$ is the surface tension and $\delta G(T)$ is the free energy density difference between liquid and crystal.
The relaxation of elastic stress is encoded in the fact that the elastic energy depends on the ratio betwen a time-scale
of nucleus formation, $f(\tau_{\mathrm N})$, and the relaxation time $\tau_{\mathrm R}$: for $\tau_{\mathrm R} \ll
f(\tau_{\mathrm N})$, $E_{\mathrm elastic}(\infty) = 0$, and the stress is relaxed. In \cite{cav05} we assumed
$f(\tau_{\mathrm N})\sim \tau_{\mathrm N}$, whereas a more accurate hypothesis is that 
$f(\tau_{\mathrm N})\sim (\log \tau_{\mathrm N})^3$ \cite{note3}, coming from the fact that the time to build a
nucleus scales as the square of the number of particles in the
nucleus. This modification makes elastic effects quantitatively even
more important. Indeed it shift the critical value of the effective
elastic coupling (which now becomes $\lambda_c=1$, see \cite{cav05}). 
On the other hand, qualitatively,  the physical picture remains unchanged. 
In the present numerical study we are interested to the case $\tau_{\mathrm R} \gg f(\tau_{\mathrm N})$, and thus we
can assume $E_{\mathrm elastic}$ constant, as in solids. This approximation is even more reasonable for
$f(\tau_{\mathrm N})=(\log \tau_{\mathrm N})^3$, since the argument of $E_{\mathrm elastic}$ becomes very small as soon as we move away from the melting 
temperature.

\begin{figure}[ht]
\begin{center}
\includegraphics[clip=true,width=8cm]{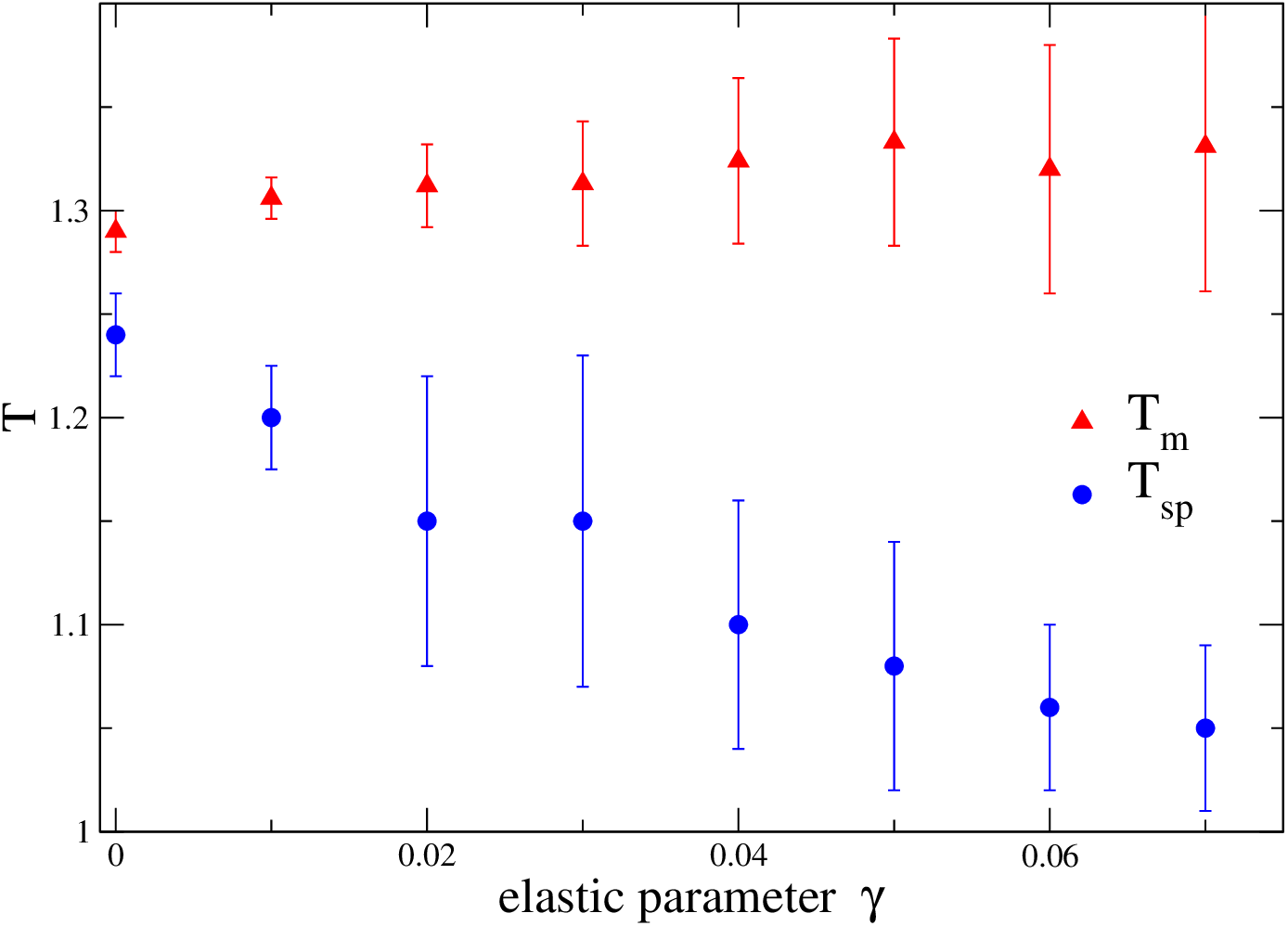}\\
\end{center}
\caption{\label{spinodal}
Kinetic spinodal $T_{sp}$ and melting temperature $T_m$ vs 
elastic parameter $\gamma$.}
\end{figure}

Nucleation with volume mismatch in an elastic solid 
has been studied in different contexts\cite{fra99,bus05}. 
In the case of an isotropic systems the elastic cost for nucleation
(or for any inhomogeneous configuration) 
 takes a simple form\cite{bus05}: 
\begin{eqnarray}
E_{elastic}=\frac{(d-1)KG}{dK+2(d-1)G}\int_V {\bigg(\epsilon_{\scriptscriptstyle{1}}^{\scriptscriptstyle 0}({\mathbf x})-\bar \epsilon_{\scriptscriptstyle{1}}^{\scriptscriptstyle 0}\bigg)}^2 d{\mathbf x}\,.
\end{eqnarray}
where $d$ is the dimension of the system (in our case $d=2$), 
$G$ ($K$) is the infinity frequency shear (bulk) modulus,
$\epsilon^{\scriptscriptstyle \circ}_{\scriptscriptstyle 1}({\bf
  x})$ is the trace of the stress free-dilation strain tensor and
$\bar \epsilon_{\scriptscriptstyle{1}}^{\scriptscriptstyle 0}$ is its
mean value. We assume that  $G$ and $K$ are the same for both phases.  
This formula  tells us that the system tries to prevent
heterogeneities. In order to implement this physics in our model we 
first rewrite the elastic energy as:
\begin{eqnarray}
E_{elastic}=A\int_V {\epsilon_{\scriptscriptstyle{1}}^{\scriptscriptstyle 0}({\mathbf x})^2 d{\mathbf x} -AV(\bar\epsilon_{\scriptscriptstyle{1}}^{\scriptscriptstyle 0}})^2\,,
\end{eqnarray}
where $A=KG/(2K+2G)$ (in $d=2$). 
The trace of the strain tensor is a 
measure of the volume change\cite{landau}, so we link it with
$m_i$ assuming a linear relationship in the spirit of Vegard's law in
alloys\cite{fra99}: 
\begin{eqnarray}
\epsilon_{\scriptscriptstyle{1}}^{\scriptscriptstyle 0}({\mathbf x}_i)=\delta\bigg(m_i-\frac{1}{2}\bigg)\,,
\end{eqnarray}
where $\delta=(\nu_C-\nu_L)/\nu_0$ and $\nu_0=(\nu_C+\nu_L)/2$.
The hamiltonian with the elastic term ($p=0$) becomes:
\begin{eqnarray}
H=H_{CTLS}+E_{elastic}=H_{CTLS}+\gamma\Bigg[\sum_{i=1}^N {m_{i}}^2-\frac{1}{N}{\bigg(\sum_{i=1}^N {m_i}\bigg)}^2\Bigg]\,,
\end{eqnarray}
where $\gamma=A\nu_0\delta^2$ is the effective elastic coupling constant.
We stress again that we are assuming that the supercooled  liquid behaves as an
elastic solid. This is true only at times shorter than the
structural relaxation time. At longer times the stresses that give
rise to the above elastic energy cost will relax. That said, we want to  
test how the elastic contribution  affects the kinetic spinodal when the
structural relaxation time is very long. 
Our simulations (Fig.~\ref{spinodal}) show that the melting 
temperature $T_m$ is approximately constant in $\gamma$. 
On the other hand, the elastic term enhance the free 
energy barrier to nucleation, and thus the spinodal temperature drastically 
decreases. As a consequence, the temperature range where the  metastable liquid is defined,
$T_{sp} < T < T_m$, gets wider and wider as the elastic coupling $\gamma$ is increased.

\section{Summary and conclusions}
Recently we have argued that viscoelasticity is the main factor determining
the existence or not of a kinetic spinodal in a supercooled liquid. 
Our arguments relied on the assumption that elasticity 
depress  nucleation, and thus shifts the kinetic spinodal.
In \cite{cav05} we performed a self-consistent determination of the nucleation time, 
taking into account viscoelastic effects. In this work we tested the
above assumption  in a microscopic model but neglecting relaxational effects. 

The supercooled liquid is strongly influenced by elastic effects as
 expected, and our 
results underline the importance of these effects. Elastic effects
strongly increase the range of metastability of the liquid phase. 
Due to technical reasons we can not increase $\gamma$ beyond $0.07$. 
This prevents us to answer the interesting question of whether the kinetic 
spinodal can cross Kauzmann temperature $T_K$, when elasticity is increased. 
Should this be the case, one expects this system to be an example of an
ideal glass former.

\section{Acknowledgments}
We thank 
C. Di Castro and J. Dyre 
for many useful comments and clarifying discussions. 

\bibliographystyle{prsty_no_etal}
\bibliography{Molveno2}

\begin{thebibliography}{1}
\bibitem{note}
For every site we define $m_i$ in this way. First we set $m_i=0$ for all sites.
  Then the sites are considered sequentially. If the spin is $\sigma_i=-1$ and
  it is sourranded by four spin whose values is $+1$, we increase $m_i$ by 1
  for every of these five sites. One can check that with this definition the
  ground state found on Ref.\cite{cavagna03} has $m_i=1$ for every site, while
  in the liquid phase $m_i$ has, on average, a small value.
\bibitem{note2}
In some configurations $m_i$ can be larger than one but this configurations are
  very rare.
\end{thebibliography}

\end{document}